\documentclass[doublecol]{epl2} 
\usepackage{graphicx}
\usepackage{amsmath,amssymb,revsymb}
\usepackage{color}
\usepackage{slashbox}

\def\cross{\times}  

\def\bfB{\mbox{\boldmath $B$}}
\def\bfu{\mbox{\boldmath $u$}}
\def\bfnabla{\mbox{\boldmath $\nabla$}}
\def\bfzero{\mbox{\boldmath $0$}}
\def\bfn{\mbox{\bf n}}
\def\bfe{\mbox{\bf e}}

\title{Effect of magnetic boundary conditions on the dynamo threshold of von K\'arm\'an swirling flows}
\author{C. Gissinger\inst{1,2}, A. Iskakov\inst{3,4}, S. Fauve\inst{1}, E. Dormy\inst{2,3}}
\shortauthor{Gissinger \etal}
\institute{
\inst{1} Laboratoire de Physique Statistique, \'Ecole Normale Sup\'erieure CNRS, 24 rue Lhomond, F-75005 Paris (France).\\
\inst{2} Laboratoire de Radioastronomie, \'Ecole Normale Sup\'erieure CNRS, 24 rue Lhomond, F-75005 Paris (France).\\
\inst{3} Institut de Physique du Globe de Paris (France).\\
\inst{4} U.C.L.A., Physics Department (U.S.A.).
}


\abstract{We study the effect of different boundary conditions on the
  kinematic dynamo threshold of von K\'arm\'an type swirling flows in a
  cylindrical geometry. Using an analytical test flow, we model different
  boundary conditions: insulating walls all over the flow, effect of sodium
  at rest on the cylinder side boundary, effect of sodium
  behind the impellers,  effect of impellers or side wall made of a high
  magnetic permeability material. We find that using high magnetic
  permeability boundary conditions decreases the dynamo threshold,  the
  minimum being achieved when they are implemented all over the flow.}

\begin{document}

\maketitle

Dynamo action, i.e., self-generation of magnetic field by the flow of an
electrically conducting fluid, is at the origin of planetary, stellar and
galactic fields \cite{moffatt}. Fluid dynamos have been observed only
recently in laboratory experiments in Karlsruhe \cite{karlsruhe} and Riga
\cite{riga} by geometrically constraining the flow lines in order to mimic
laminar flows that were known analytically for their dynamo efficiency
\cite{robpom}. More recently, the VKS experiment  displayed self-generation
in a less constrained geometry, e.g. a von K\'arm\'an swirling flow
generated between two counter-rotating impellers in a cylinder
\cite{monchaux07}.   
However, until now, dynamo action in the VKS geometry has been found only
when the impellers are made of soft iron. It is thus of primary importance
to understand how the dynamo problem is modified by the presence of
magnetic material at the flow boundaries. We address this problem here
using a kinematic dynamo code in a cylindrical geometry.  Two important
approximations are made to simplify the study. First,  an analytic test
flow that mimics the geometry of the mean flow of the VKS experiment is
considered. Second, the magnetic boundary conditions are taken in the limit
of infinite magnetic permeability of the boundaries compared to
the one of the fluid. This seems a reasonable approximation for soft iron
compared to liquid sodium. Our main result is that the critical magnetic
Reynolds number, $Rm_c$, for dynamo generation is significantly decreased with
boundaries of high magnetic permeability all over the flow.

The VKS experimental set-up is sketched in Figure~1.   A turbulent von
K\'arm\'an flow of liquid sodium  is generated by two counter-rotating
impellers (rotation frequencies $F_1$ and $F_2$). The impellers are made of
iron disks of radius $154$ mm, fitted with $8$ iron blades of height
$41.2$~mm, 
and are placed $371$~mm apart in an inner cylinder of radius $206$ mm
and length $524$~mm. It is surrounded by sodium at rest in another
concentric cylindrical vessel, $578$ mm in inner diameter. This has been
shown to decrease the dynamo threshold in kinematic computations based on
the mean flow velocity \cite{marie03}.   
When the impellers are operated at equal and opposite rotation rates $F$, a statistically stationary magnetic field 
is generated above a magnetic Reynolds number $R_m \sim 30$~\cite{monchaux07}. The large scale field involves an azimuthal 
component and a poloidal one which is dominated by an axial dipole.   This
geometry has been understood with a simple  $\alpha-\omega$
dynamo model  \cite{petrelis07} by taking into account the helical nature of the
flow that is ejected by the centrifugal force close to each impeller
between successive blades. Relying on  the mean flow alone to
compute the kinematic dynamo, smoothes out these non axisymmetric velocity
fluctuations  and thus cannot generate an axisymmetric field according to
Cowling theorem. A non axisymmetric field is obtained, dominated by
an equatorial dipole \cite{marie03,bourgoin04}.  

\begin{figure}[!htb]
\centerline{\includegraphics[width=9cm]{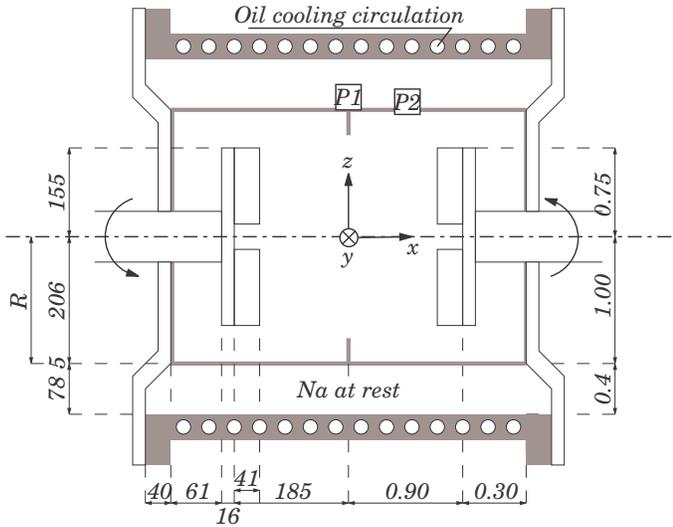}}

\caption{Sketch of the VKS experiment \cite{monchaux07}. The inner and outer cylinders are made of copper (in gray). The dimension are given in millimeter (left) and normalized by the radius of the inner cylinder (right).} 
\label{vks}
\end{figure}

When the disks are counter-rotating at the same frequency, the structure of
the mean flow (averaged in time) has the following characteristics: the
fluid is ejected radially from the disks by the centrifugal force and loops
back towards the axis in the mid-plane between the impellers. A strong
differential rotation is superimposed on this poloidal flow, which
generates a high shear in the mid-plane. 
We approximate the experimental configuration with impellers of radius $R$ 
(i.e. extending up to the inner cylinder boundary).
We use cylindrical coordinates $(s , \phi, z)$.
In this cylindrical domain
($[0,1]\times[0, 2\pi]\times[-1,1]$), the flow is well described
by the following analytical expression 
\begin{subequations} 
\begin{equation}
\bfu = \bfnabla \times \left( \psi \, \bfe_\phi \right) + s \omega \, \bfe_\phi
\, ,
\end{equation} 
with angular velocity $\omega$  and recirculation $\psi$ respectively given by
\begin{equation}
\omega = 4 \, \varepsilon \, \left(1-s\right) \,
\sin\left(\tfrac{\pi}{2}z\right) \, ,
\end{equation} 
\begin{equation}
\psi = {s\over 2}{\left(1-s\right)}^2\left(1+2s\right)\, \sin\left(\pi
z\right) \, .
\end{equation} 
\label{flow}
\end{subequations} 
This flow has been shown to generate a similar kinematic dynamo as the one
computed using the experimentally measured mean flow \cite{marie06}.  
In the above expression, $\varepsilon$ is a parameter
controlling the ratio between the poloidal and toroidal components
of the flow.  We take $\varepsilon=0.7259$ as in previous numerical studies
using this flow, in order to minimize the critical magnetic Reynolds 
number for dynamo threshold \cite{marie06,ravelet05}.

Although we are aware that the experimentally observed dynamo cannot be captured with a kinematic calculation using the mean flow alone, we use this simple model here in order to study the effect of the magnetic boundary conditions on the dynamo threshold.  
We hope that the qualitative behaviors will be unchanged with other dynamo modes in the presence of 
a turbulent flow.

We perform direct numerical simulations of kinematic dynamos, solving the
induction equation governing the evolution of the solenoidal magnetic field
$\bf B$  
\begin{equation}
\frac{\partial \bfB}{\partial t} = Rm\,\bfnabla \times \left(
\bfu \times \bfB \right) + \Delta \bfB \, ,
\label{induction}
\end{equation}
written in dimensionless form, using the diffusive timescale.
The magnetic Reynolds number $Rm$ is defined as
$Rm=\mu _0 \sigma R U_{max}$,
where $R$ is the cylindrical radius of the domain of the flow defined by 
(1a-c) and $U_{max}$ is the peak velocity of the flow.\\

Equation (\ref{induction}) with the flow given by (\ref{flow}a-c) is
solved using a finite volume code adapted from \cite{teyssier}.  To
circumvent the severe CFL restriction  induced by
cylindrical coordinates, we ensure numerical stability
near the axis using a low pass Fourier filter in the
$\phi$--direction. Also a centered second order scheme has been
preferred fo to an up-wind scheme to discretize the inductive term,
as resistive 
effects are here important
enough to regularise the solution.  As in \cite{teyssier}, we ensure
that $\bfnabla \cdot \bfB=0$ is exactly satisfied using a constraint
transport algorithm.  The finite volume solver is fully
three-dimensional. We have not used the decoupling of Fourier modes in
the $\phi$--direction.
The initial magnetic field obviously needs to satisfy the divergence-free 
constraint as well as the boundary conditions. In practice, 
an arbitrary divergent free field is initialized away from all boundaries.

We investigate several types of magnetic boundary conditions.
The classical approach is to use insulating boundaries, matching the internal 
magnetic field to the vacuum magnetic potential.
The continuity of the magnetic field 
results in a non-local set of boundary conditions, 
which can be expressed via a ``Neumann to Dirichlet'' operator.
We rely here on such an approach, using the boundary element formalism, 
as introduced in \cite{iskakov}. 

We investigate the effects introduced by using ferromagnetic
boundaries.  This boundary condition can be expressed in a local form
in the limit of infinite permeability.  Jump conditions at a boundary
between media of different magnetic permeability are well established
\begin{equation}
\bfB \cdot \bfn |_{2}=\bfB \cdot \bfn |_{1} \, , 
\label{CL1}
\end{equation}
\begin{equation}
\bfB \cross \bfn |_{2}={\mu _2\over \mu _1} \bfB \cross \bfn |_{1} \, ,
\label{CL2}
\end{equation}
where subscripts $1$ and $2$ denote the two different regions and $\bfn$ is
normal to the boundary.
Ferromagnetic disks yield $\mu \gg \mu_0 $, one can therefore reasonably
approximate these jump relations, by using boundary conditions on the fluid
side of the form
\begin{equation}
\bfB \cross \bfn = \bfzero \, .
\label{CL3}
\end{equation}
This set of boundary conditions trivially implies that normal currents
vanish. 

We will consider the effect of iron disks in two different ways. The
first, and probably simpler, approach is to assume that the field
is normal to the disks, namely
\begin{equation}
\bfB \cross \bfe_z = \bfzero \, ,
\end{equation}
on the top and bottom of the cylinder.
This boundary condition is well known in magnetohydrodynamics,
in particular in the astrophysical community, and is sufficient
to close the system of equations we investigate.
Another, and more subtle approach, allows to take into account 
the effect of the blades on the disks. Assuming that the field 
becomes normal to radial blades as it approaches the end of the cylinder yields
\begin{equation}
\bfB \cross \bfe_\phi = \bfzero \, .
\end{equation}
This is an extremely simplified approach which does not take into account
each individual blade, but accounts for their average effect on the large 
scale field.
This set of boundary conditions is far less common, and deserves some
care to ensure it provides the required constraints on the field.
The solenoidal nature of magnetic
field $(\bfnabla \cdot \bfB=0)$, together with the fact that $B_s$ is identically
zero on the boundary, then imply
\begin{equation}
{\partial B_\phi \over \partial \phi} = - s \, {\partial B_z \over \partial
  z}\, .
\label{la divergence}
\end{equation} 
This yields the solvability condition
\begin{equation}
\oint_0^{2\pi} {\partial B_z \over \partial z} \,d\phi = 0 \, .
\label{solvability}
\end{equation}
As the flow we consider here is axisymmetric, modes in the
$\phi$--direction decouple, 
and we know from Cowling's theorem that the magnetic eigenmode cannot be 
axisymmetric. 
The solvability condition is therefore
obviously satisfied (this would not be the case for a non-linear simulation).

For non axisymmetric modes, one can write
\begin{equation}
\oint_0^{2\pi} B_\phi \, d\phi=0 \, , 
\label{Bphi}
\end{equation}
which together with (\ref{la divergence}) determines the $B_\phi$
field completely. The numerical implementation is two fold, we first
compute
\begin{equation}
B'_\phi(\phi)=-s\,\oint_{0} ^{\phi} {\partial B_z \over \partial
z}d\phi \, ,
\label{B'phi}
\end{equation}
up to an arbitrary constant. Then we correct this function setting
by the constant to meet (\ref{Bphi}).

This implementation is mathematically consistent and maintains the
divergence free property of the magnetic field. We shall however
stress again that this very idealized condition can only be used
here because the modes decouple in the azimuthal direction and 
the axisymmetric mode cannot be unstable.


The convergence of the numerical implementation has been carefully
validated comparing simulations at different resolutions.  We report
results obtained with a resolution of $200 \times 200 \times 256$
points.  When an additional domain of sodium at rest is included, we
use the same resolution in the inner flow domain defined by (1a-c), which
leads for the full domain to $240$ points in the radial direction and
$250$ points in the 
$z$--direction to maintain a uniform accuracy.

\begin{figure}[!htb]
\centerline{\includegraphics[width=9cm]{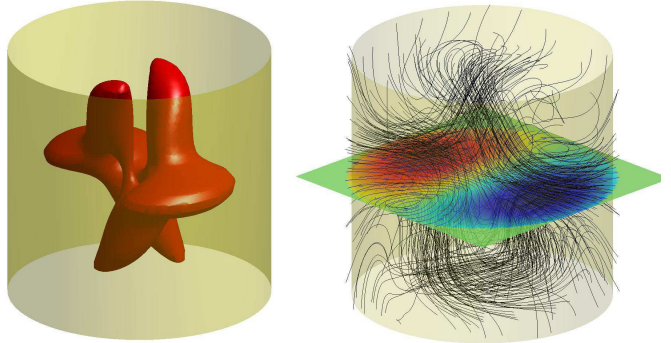}}
\caption{The magnetic eigenmode obtained with potential boundary conditions. The
  rotation axis is vertical. 
An isosurface of the magnetic energy ($25\%$ of the
maximum value) is represented on the left.
Magnetic field lines are plotted on the right.} 
\label{fig mode neutre}
\end{figure}

As in previous studies, all simulations yield magnetic eigenmodes
with an $m=1$ azimuthal symmetry. The structure of this eigenmode 
essentially corresponds to an equatorial dipole. 
This mode is represented on Figure~\ref{fig mode neutre}.

\begin{figure}[!htb]
\centerline{\includegraphics[width=11cm]{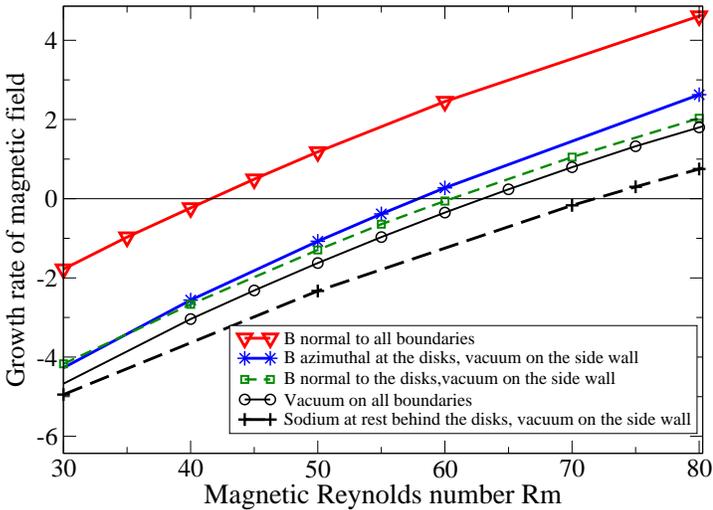}}
\caption{Growth rate of the magnetic field as a function of $Rm$ for
  different types of boundary conditions. We can see that replacing
  sodium at rest behind the disks by vacuum leads to a reduction of the
  dynamo threshold. Moreover, ferromagnetic conditions lead to
  additional reduction of threshold (about $10 \%$ for azimuthal
  conditions). Note that using high magnetic permeability boundary 
  conditions on the disks but
  also on the side yields the best configuration, with an onset at
  $Rm_c=41.5$. }
\label{fig result1}
\end{figure}

Previous numerical studies \cite{laguerre06,stefani06} compared the
threshold values in configurations including and excluding the effect
of fluid behind the disks. 
We reproduce here a similar behaviour of the threshold:
using insulating boundary conditions directly on the disks yields
a threshold value $Rm_c=63$. 
This last value is in good agreement with 
\cite{laguerre06}.
Adding a layer of thickness $0.25$ of fluid at rest between each disk 
and insulating boundaries
increases the dynamo onset up to $Rm_c=72$.
This is illustrated on Figure~\ref{fig result1}.
Simulations, not discussed here, show 
that if the sodium behind the disks is not at rest, the threshold is 
even further increased.

When we consider the case of ferromagnetic
disks, the threshold for dynamo action is $Rm_c=60.5$ for a magnetic
field normal to the disks and $Rm_c=58$ for a magnetic field in the
$\phi$--direction. These two thresholds are very close and lower than
in the case of a potential field (see also Figure~\ref{fig result1}).
We thus observe that the high permeability boundary condition on 
the disks appears to do more than just screening the inhibition of 
electromagnetic induction due to the flow behind the disks. 

Implementation of a cylindrical layer of sodium at rest on the side 
can easily be included in the simulations. Since all electrical quantities 
are continuous, and in particular 
there is no jump in the conductivity, this does not involve any extra 
boundary condition. In some sense, the model assumes a continuous fluid 
whose velocity field goes to zero outside a given domain.
Table~\ref{tableau} reports critical parameters
calculated for different combinations of boundary conditions on the
side and on the disks. The sodium at rest leads to an important
reduction of dynamo thresholds when it is added on the side. It is
known \cite{ravelet05,laguerre06,stefani06} that this effect
increases with the width of the layer, but saturates relatively rapidly. In
all the simulations presented, the width of the layer of sodium at
rest on the side is set to $0.4$ in order to study a configuration close to the VKS
experiment. This positive impact on the onset appears to be independent of 
the boundary conditions on the disk.

Finally, observing the strong effect produced by ferromagnetic disks,
it is tempting to investigate the effect of high permeability material
for the entire vessel. We thus perform simulations for which the magnetic field
is normal to all boundaries (see Figure~\ref{fig result1}). 
This configuration appears to be the most efficient and it
leads to a critical magnetic Reynolds number $Rm_c=41.5$. It is clear from
Table~\ref{tableau} that in the presence of sodium at rest,
ferromagnetic boundaries still yield a strong reduction of the onset,
despite the fact that the boundary is now remote from the flow domain.  
No significant modification of the global magnetic
structure can be observed in our simulations when we compare
ferromagnetic and vacuum boundary conditions. The magnetic eigenmodes 
in the bulk are similar with both sets of boundary conditions. 

\begin{table}
\begin{tabular}{|l|c|c|c|c|}
  \hline \backslashbox{disks}{side} &$Na\,\&\, B_n$ & $B_n$
  & $Na\,\&\,-{\bf\nabla}\Phi$ &$-{\bf\nabla}\Phi$ \\ \hline $B_{\phi}$ & 39 & 41 & 45& 58
  \\ \hline $B_n$ & 40 &41.5& 45.5 & 60.5 \\ \hline
  $-{\bf\nabla}{\Phi}$ & 44 &43.5& 48.5 & 63 \\ \hline $Na\,\&\,B_{\phi}$
  & 46 & 47 & 53 & 71 \\ \hline $Na\,\&\,B_n$ & 47 &47.5& 53.8 & 71.5
  \\ \hline $Na\,\&\,-{\bf\nabla}{\Phi}$ & 47.5& 48 & 54.7 & 72 \\
 \hline
\end{tabular}
\caption{Dynamo thresholds for different types of boundary conditions
  on the disks and on the sides of the cylinder. $B_n$ and $B_{\phi}$
  denote ferromagnetic boundaries (respectively normal to the disks and normal to the blades), 
$-{\bf\nabla}\Phi$ denotes the vacuum condition and $Na$ indicates the presence of a layer of 
sodium at rest between the flow and the boundary. 
For the disks, cases $Na\,\&\,B_{\phi}$, $Na\,\&\,B_n$ or $Na\,\&\,-{\bf\nabla}{\Phi}$ correspond 
to the presence of a layer of sodium at rest, outside which the
  relevant boundary condition is applied. 
The sodium lies behind the disks, and the boundary conditions are implemented on the lids.}
\label{tableau}
\end{table}

The effect of the electrical conductivity of the boundaries on the dynamo threshold
has been studied since a long time \cite{bullard}. It has been shown that the addition
of a layer of electrically conducting fluid at rest around the flow can either decrease or
increase the threshold \cite{kaiser99}. This has also been observed in the case
of  von K\'arm\'an flows, depending on the location of the layer at rest
\cite{stefani06,laguerre06}. Thus, general rules for the dependence of the dynamo 
threshold on the electrical conductivity of the boundary  do not seem to exist. 
Such may not be the case with the magnetic permeability of the bounding domain.
Previous studies on the Ponomarenko dynamo with high magnetic permeability 
boundaries have displayed a decrease of the threshold \cite{marty95,avalos03}.
A similar improvement has been observed in the case of convectively driven dynamos
\cite{Morin}. In the present study, we find that boundaries with a high magnetic permeability
always decrease the dynamo threshold whatever their location. 

Two important aspects of the VKS experiment are not taken into account in the present 
study. First, the magnetization of iron can lead to an additional amplifying factor
of the dynamo as discussed in \cite{petrelis07}. However, the coercitive field of pure 
iron being much smaller than the fields generated by the dynamo, the iron disks do not impose any permanent magnetization.  Reversals of the generated magnetic field are indeed observed \cite{berhanu07}. 
Second, the geometry of the magnetic field generated in the VKS experiment differs from the one 
computed from the axisymmetric mean flow. Non axisymmetric velocity fluctuations generate a 
poloidal field with a dominant axial dipole, together with a strong azimuthal component. 
Thus the poloidal field is roughly normal to the disks, whereas the toroidal field is normal to the blades. 
It would therefore be of interest to check whether the dynamo threshold can be reached by using only 
ferromagnetic disks or blades. These two 
cases would correspond respectively to our boundary conditions (6) and (7).

Another interesting set of modifications suggested by this numerical study, would be 
to replace the copper side wall by a ferromagnetic one in the VKS experiment.
If the experimentally realised dynamo mode behaves similarly to the simulations, we 
expect the threshold to be even lower in such configuration than that associated 
with only ferromagnetic disks and blades.

\acknowledgements


\begin{thebibliography}{0}

\bibitem{moffatt}
See for instance, 
\Name{Moffatt H.K.}
  \Book{Magnetic field generation in electrically conducting fluids}
  \Publ{Cambridge University Press, Cambridge}
  \Year{1978};
\Name{Dormy E. \and Soward A.M.}
  \Book{Mathematical Aspects of Natural Dynamos}
  \Publ{CRC-press}
  \Year{2007}.
  
  \bibitem{karlsruhe}
  \Name{Stieglitz R. \and M\"uller U.}
  \REVIEW{Phys. Fluids}{13}{2001}{561}.

 \bibitem{riga}
  \Name{Gailitis A. \etal}
  \REVIEW{Phys. Rev. Lett.}{86}{2001}{3024}.
 
 \bibitem{robpom} 
\Name{Roberts G. O.}
\REVIEW{Phil. Trans. Roy. Soc. London A}{271}{1972}{411};
\Name{Ponomarenko Yu. B.}
\REVIEW{J. Appl. Mech. Tech. Phys.}{14}{1973}{775}.

 \bibitem{monchaux07}
  \Name{Monchaux R. \etal}  
  \REVIEW{Phys. Rev. Lett.}{98}{2007}{044502}.

\bibitem{marie03} 
\Name{Mari\'e L. \etal}
 \REVIEW{Eur. Phys. J. B}{33}{2003}{469}.
 
 \bibitem{petrelis07}
  \Name{P\'etr\'elis F., Mordant  N. \and Fauve S.}  
  \REVIEW{Geophys.Astrophys. Fluid Dyn.}{101}{2007}{289}.

 \bibitem{bourgoin04} 
\Name{Bourgoin M. \etal}
  \REVIEW{Phys. Fluids}{16}{2004}{2529}.

\bibitem{marie06} 
 \Name{Mari\'e L., Normand C.  \and Daviaud F.}
  \REVIEW{Phys. Fluids}{18}{2006}{017102}.
  
   \bibitem{ravelet05}
  \Name{Ravelet F. \etal}
  \REVIEW{Phys. Fluids}{17}{2005}{117104}.

 \bibitem{teyssier} 
   \Name{Teyssier R., Fromang S. \and Dormy E.}
   \REVIEW{J. Comp. Phys.}{218}{2006}{44}.
 
 \bibitem{iskakov} 
   \Name{Iskakov A., Descombes S., \and Dormy E.}
  \REVIEW{J. Comp. Phys.}{197}{2004}{540};
   \Name{Iskakov A. \and Dormy E.}
   \REVIEW{Geophys. Astrophys. Fluid Dyn.}{99}{2005}{481}.
  
  \bibitem{stefani06} 
\Name{Stefani F. \etal}
\REVIEW{Eur. J. Mech. B}{25}{2006}{894}.

\bibitem{laguerre06} 
 \Name{Laguerre R. \etal}  
 \REVIEW{CRAS M\'ecanique}{334}{2006}{p.593-598}.


\bibitem{bullard} 
   \Name{Bullard E. C. \and Gubbins D.}
   \REVIEW{Geophys. Astrophys. Fluid Dyn.}{8}{1977}{43}.

\bibitem{kaiser99}
\Name{Kaiser R. \and Tilgner A.}
\REVIEW{Phys. Rev. E}{60}{1999}{2949}.

\bibitem{marty95}
\Name{Marty P., Ajakh A.  \and Thess A.}
\REVIEW{Magnetohydrodynamics}{30}{1995}{474}.

\bibitem{avalos03}
\Name{Avalos-Zuniga R., Plunian F. \and Gailitis A.}
\REVIEW{Phys. Rev. E}{68}{2003}{066307}.

\bibitem{Morin} 
 \Name{Morin V.}  
 \REVIEW{PhD Thesis, University Paris VII}{}{2005}{ Appendix C.}


\bibitem{berhanu07} 
 \Name{Berhanu M. \etal}  
 \REVIEW{Europhys. Lett.}{98}{2007}{59001}.


 \end{thebibliography}
\end{document}